\documentclass[fleqn,usenatbib]{mnras}

\usepackage{newtxtext,newtxmath}

\usepackage[T1]{fontenc}
\usepackage{ae,aecompl}

\usepackage{graphicx}	
\usepackage{amsmath}	
\usepackage{amssymb}	

\newcommand{\logMstarT}{$10.81 \pm 0.04$}
\newcommand{\logphistarT}{$-3.43 \pm 0.02 $}
\newcommand{\alphaT}{$-0.62 \pm 0.09$}
\newcommand{\logMstarM}{$10.89 \pm 0.06$}
\newcommand{\logphistarM}{$-4.30 \pm 0.03 $}
\newcommand{\alphaM}{$-0.55 \pm 0.08$}
\newcommand{\logphistarThigh}{$-3.53 \pm 0.02 $}
\newcommand{\logphistarTmed}{$-3.43 \pm 0.02 $}
\newcommand{\logphistarTlow}{$-3.36 \pm 0.02 $}

\title[Transfer learning for merger detection]{Using transfer learning to detect galaxy mergers}

\author[S. Ackermann et al.]
{Sandro Ackermann,$^{1}$
Kevin Schawinski,$^{1}$
Ce Zhang,$^{2}$ 
Anna K. Weigel$^{1}$
and 
\newauthor M. Dennis Turp$^{1}$
\\
$^{1}$Institute for Particle Physics and Astrophysics, Department of Physics, ETH Zurich, Wolfgang-Pauli-Strasse 27,\\
CH-8093, Z\"{u}rich, Switzerland\\
$^{2}$Systems Group, Department of Computer Science, ETH Zurich, Universit\"{a}tstrasse 6, CH-8006, Z\"{u}rich, Switzerland\\
}
\date{Accepted XXX. Received YYY; in original form ZZZ}

\pubyear{2017}

\begin{document}
\label{firstpage}
\pagerange{\pageref{firstpage}--\pageref{lastpage}}
\maketitle

\begin{abstract}
We investigate the use of deep convolutional neural networks (deep CNNs) for automatic visual detection of galaxy mergers. Moreover, we investigate the use of transfer learning in conjunction with CNNs, by retraining networks first trained on pictures of everyday objects. We test the hypothesis that transfer learning is useful for improving classification performance for small training sets. This would make transfer learning useful for finding rare objects in astronomical imaging datasets.
We find that these deep learning methods perform significantly better than current state-of-the-art merger detection methods based on nonparametric systems like CAS and GM$_{20}$. Our method is end-to-end and robust to image noise and distortions; it can be applied directly without image preprocessing. We also find that transfer learning can act as a regulariser in some cases, leading to better overall classification accuracy ($p = 0.02$). Transfer learning on our full training set leads to a lowered error rate from 0.038\textpm 1 down to 0.032\textpm 1, a relative improvement of 15\%.
Finally, we perform a basic sanity-check by creating a merger sample with our method, and comparing with an already existing, manually created merger catalogue in terms of colour-mass distribution and stellar mass function.
\end{abstract}




\section{Introduction}

Galaxy mergers are an important driver of the mass assembly and transformation of massive galaxies and the triggering of quasars \citep{1972ApJ...178..623T, 1998A&A...331L...1S, 1988ApJ...325...74S, 1996ApJ...464..641M, 2003ApJS..149..289B, 2005MNRAS.361..776S, 2006ApJS..163....1H, 2010Sci...328..600T, 2012ApJ...758L..39T, 2017ApJ...845..145W}. Several different methods have been previously used to detect galaxy mergers in observational data:

One way to detect mergers is the \textit{close-pairs} method (e.g. \cite{Lin2004} or \cite{Woods2007}). Here, luminosity
peaks are algorithmically identified. Each one of these peaks
is then considered as a target for collecting spectroscopic data. Close pairs
of such spectroscopic targets in the image plane are considered as potential
mergers. Redshift measurements can then be used to confirm that two candidates are close enough to be gravitationally
interacting in a significant way. One problem with the close pairs method is imprecise radial distance
measurements due to peculiar velocities. Furthermore, detecting the
spectrographic targets requires a heuristic algorithm, which may be hard to
hand-engineer, considering the diversity in the morphologies of mergers. It might
even be impossible, for some merger images, to detect two separate luminosity peaks, as not every merger exhibits two distinct luminosity peaks.

Another approach purely relies on imaging data. Here, a handful of hand-crafted feature detectors are used for classification. Examples for this approach are the CAS features (concentration, asymmetry, clumpiness) by \cite{Conselice2003}
or GM$_{20}$ (Gini coefficient and the second-order moment of the brightest 20\% of the galaxy's flux) by \cite{Lotz2004} and combinations and variants of those systems
\citep{Cotini2013, Hoyos2011, Goulding2017}. Those \textit{nonparametric} systems prescribe, in an algorithmic manner, how each feature should be extracted from a galaxy image; e.g. asymmetry from CAS is
defined as the normalised residuals of comparing a flipped version of the galaxy image to the original image.
Determining the individual values for each of the $n$ features allows the classification algorithm to distinguish between mergers and non-interacting galaxies, depending on where the analysed galaxy image lies in the $n$-dimensional feature space.
A known problem with these nonparametric approaches using hand-crafted feature detectors is the difficulty of capturing the full diversity of merger appearances, and the varying sensitivity of detection to each stage of the merger process, see the simulation studies by \cite{Lotz2008} and \cite{Lotz2010}.

Until now, the accuracy of manual classification by human experts cannot be reached by automatic methods. The Galaxy Zoo (GZ)\footnote{\url{http://zoo1.galaxyzoo.org/}} project by \cite{Lintott2008, Lintott2010} achieved visual morphological classification of around one million galaxies from the Sloan Digital Sky Survey (SDSS) by crowdsourcing the classification task out to citizen scientists on an online platform. Classifying mergers versus non-interacting galaxies was part of the first Galaxy Zoo run, and we will use those classifications later in this paper as our ground truth classifications to train our own classifier.

The problem of algorithmically categorising images into different classes is not a problem that is specific to merger detection in astronomy. Image classification is one of the main problems of the computer vision and machine learning community. They have developed, over the years, a wealth of methods to solve image classification. Recently, with the advent of large labeled datasets and cheap computational resources, Convolutional Neural Networks (CNNs) have achieved a performance level that represents a significant improvement over more traditional computer vision methods \citep{Krizhevsky2017}. Today, the best CNN architectures can rival or even surpass the performance of human classifiers on some datasets \citep{He2015}. Deep CNNs have already been used successfully for regression and classification tasks with imaging data of galaxies \citep{Dieleman2015, Hoyle2016}. It seems prudent to test how these state-of-the-art CNN architectures fare in our specific task of detecting mergers; this will be the main focus of this paper.

We would like to emphasise that, when using deep CNNs, there are no hand-crafted feature detectors involved. The salient features for the specific classification task are discovered by the neural network automatically during training. Training is the process of tuning the free parameters of a CNN to a given task with an optimisation algorithm. This removes the human element of coming up with meaningful feature descriptors in the first place. This property of a machine learning method is called \textit{feature learning}. To our best knowledge, this is the first use of feature learning for automatic visual galaxy merger detection. Essentially, classification using CNNs is an end-to-end method that can be applied directly to the raw pixel values, without preprocessing or dimensionality reduction. This also means that CNNs are very robust to noise or image defects, as long as they are already present in the \textit{training} data.

In this paper, we will investigate the use of CNNs to potentially achieve improvements over the previous state-of-the-art in automatic galaxy merger detection.

\section{Method}

\subsection{Deep Convolutional Neural Networks}

Artificial neural networks (ANNs) have been a focus of artificial intelligence research for more than half a century \citep{McCulloch1943, Hebb1949}.

One way to motivate the use of artificial neural networks for artificial intelligence tasks is that their biological inspiration, neural networks in animals, are tasked with information processing, for example visual processing in the visual cortex. There are some surprising parallels between the visual cortex and convolutional neural networks trained on natural images, like e.g. the emergence of Gabor-like filters in the first layer of processing \citep{Marelja1980, Daugman1985, Jones1987, Krizhevsky2017, Cichy2016}.

Another, mathematically more rigorous way to look at artificial neural networks is to interpret them as universal approximators: Neural networks of sufficient size can, in theory, approximate any input-output mapping that is desired \citep{Cybenko1989, Hornik1991}.

Artificial neural networks generally consist of units (the artificial neurons) that are connected in a directed graph. We are basing the following notation loosely on \cite{Goodfellow2016}. Each unit is assigned an activation $\tilde{y} \in \mathbb{R}$, which is based on adding up all the incoming activations, originating from the $n$ precursor units $\tilde{\textbf{x}} \in \mathbb{R}^n$, weighted by the particular weight that defines the strength of the connection of each edge $\textbf{w} \in \mathbb{R}^n$. An activation function $\phi(\cdot)$ is then used to compute the activation $\tilde{y}$ of the unit from the inputs $\tilde{\textbf{x}}$ and the weights $\textbf{w}$:

\begin{equation}
\tilde{y} = \phi(\tilde{\textbf{x}}^T \textbf{w})
\end{equation}

In our case, we are interested in acyclical graphs, and we are restricting ourselves to graphs that have "layers", i.e. an ANN that consists entirely of groups (or layers) of units that only get inputs from the activations of a precursor group of units. The first layer is the input layer, and the activations of these units gets set to the input values $\textbf{x} \in \mathbb{R}^k$. The last layer is the output layer, and the activations of these units, after calculating them from propagating the input activations through the whole network, represents the output of the neural network $\textbf{y} \in \mathbb{R}^l$. The free parameters of the neural network, given by the weights of all connections $\mathbf{\theta}$, will be used to compute the output $\textbf{y}$ from the input $\textbf{x}$, so essentially we are learning a function from $\mathbb{R}^k \mapsto \mathbb{R}^l$ that is parametrized by the weights $\mathbf{\theta}$:

\begin{equation}
\textbf{y} = f(\textbf{x} \mid \mathbf{\theta})
\end{equation}

Each layer $n \in \{1, 2, \dots, s \}$ can be seen as a function $f_n(\cdot)$, mapping from one intermediate vector space into another. Evaluating the whole network is then just the process of composing the individual layers together, $f(\textbf{x}) = f_s \circ \ldots \circ f_2 \circ f_1(\textbf{x})$. Layers that are not the input or output layer are called "hidden layers". The use of many hidden layers is what gave deep neural networks (DNNs) their name.

By changing the weights of our neural network $\mathbf{\theta}$, we can change the function $f(\cdot \mid \mathbf{\theta})$. \textit{Training} is the process of finding the optimal weights with respect to a loss function $C_T(\cdot)$, so that the neural network function $f(\cdot \mid \mathbf{\theta})$ approaches a desired, ground truth function $f^*(\cdot)$  as closely as possible on a given \textit{training set}; we want to find $\hat{\theta} = \arg\min_{\mathbf{\theta}} C_T(\mathbf{\theta})$. A training set $T$ is a set of tuples $\left(\textbf{x}_i, \textbf{y}^*_i\right)$ which were sampled from the ground truth function; $\textbf{y}^*_i = f^*(\textbf{x}_i)$.

Given that the loss landscape in our case is generally not convex, we are resorting to gradient based methods (gradient descent and related algorithms) to do this optimisation step to arrive at $\hat{\theta}$. In theory, gradient descent can converge to any local minimum, and we have no guarantee of finding a solution that is at, or close to the global minimum. However, this does not seem to pose a very significant problem with currently employed neural network architectures \citep{Goodfellow2014}. Please note that there are alternative algorithms for non-convex optimisation, which handle local minima more gracefully, e.g. evolutionary algorithms for optimisation.

Having finished the training, we can use our function $f(\cdot \mid \mathbf{\hat{\theta}})$ as an approximation of the ground truth function $f^*(\cdot)$. This second step with fixed weights is called \textit{inference}.

A specific type of layer is the convolutional layer: Inspiration from the discovery of receptive fields of neurons in the visual cortex, and taking advantage of the translational symmetry of natural images lead researchers to experiment with this type of layer for solving visual processing tasks with ANNs \citep{Fukushima1980, Lecun1998, Krizhevsky2017}. In a convolutional layer, each unit only receives input (i.e. has nonzero weights) from a local image patch of the precursor layer (\textit{receptive field}). Many units (they constitute a so called \textit{feature map}) in the convolutional layer share the same weight matrix, apart from two-dimensional translations in the image plane, so that each unit has its own receptive field. This is essentially the same as convolving the original layer with a convolution kernel corresponding to the weight matrix, and then using the activation function $\phi(\cdot)$ on each pixel. Doing this convolution for multiple different kernels yields multiple feature maps. The activations of those feature maps are essentially the activations of the convolutional layer and are the input to the next layer.

Convolutional layers exhibit translational invariance and they dramatically reduce the number of free, learnable parameters due to weight sharing and limited receptive fields. This speeds up the training process and acts as a strong regularizer. Convolutional layers are used extensively in todays state-of-the-art DNNs for visual processing (see the seminal work by \cite{Krizhevsky2017}) and other tasks. These types of DNN architectures are called deep convolutional neural networks, or deep CNNs.

We will be using the \textsc{Xception} CNN architecture by \cite{Chollet2016} for this paper.

\subsection{Transfer Learning}

The network architectures that are used in deep learning have a very high dimensional free parameter space, i.e. they have many different weights that need to be tuned during training. When trained on a limited number of samples (small training set), this gives the network enough capacity to fit to the noise or specific properties of that training set. This leads to a good accuracy on the training set, but it does not generalize well to new data. This phenomenon is called \textit{overfitting}. \textit{Regularisation} is the attempt to reduce overfitting with various methods.

Overfitting becomes especially problematic with increasingly smaller training set sizes and increasingly large (large in the sense of many free parameters) neural networks. In our case, we are attempting to classify rare astronomical objects (small training set size) with SOTA (state-of-the-art) classifier CNNs (large neural networks). To achieve good general classification performance, we need to use regularisers to combat overfitting.

\begin{figure}
    \centering
	\includegraphics[width = 0.45 \columnwidth]{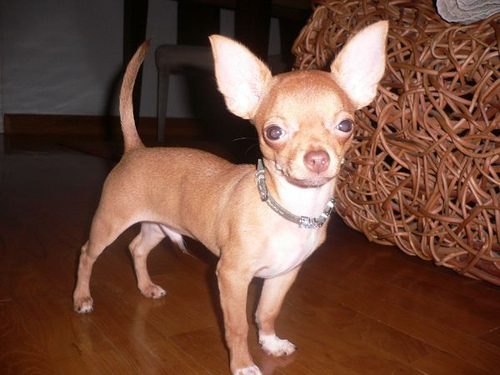}
    \includegraphics[width = 0.45 \columnwidth]{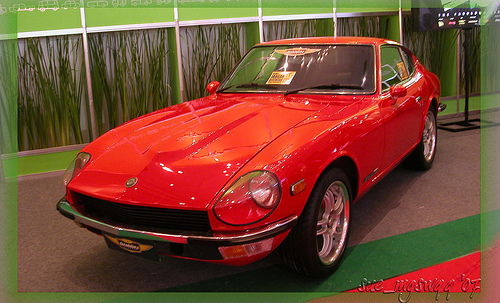}
    \includegraphics[width = 0.45 \columnwidth]{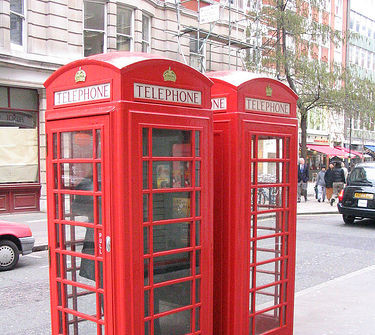}
    \includegraphics[width = 0.45 \columnwidth]{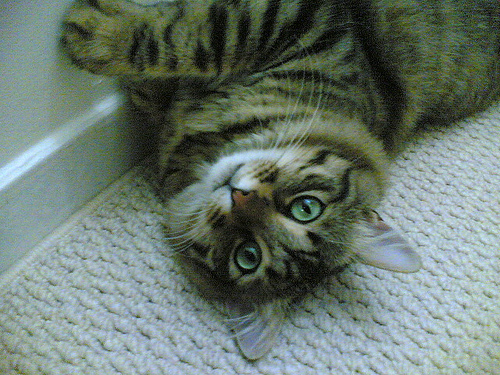}
    \caption{Sample images from the \textsc{ImageNet} dataset, belonging to categories with labels \textit{Chihuahua}, \textit{Sports car}, \textit{Telephone booth} and \textit{Tiger cat}.}
    \label{fig:imagenet_sample_images}
\end{figure}

In this paper, one objective is to show the following: We can pre-train a SOTA classifier CNN on a the \textsc{ImageNet} dataset \citep{JiaDeng2009}, which is a big dataset of a few million natural images from thousands of categories, containing categories like cars and dogs and cats and many others. A selection of some sample images including their class labels can be seen in figure \ref{fig:imagenet_sample_images}. We then use the CNN with these pre-trained weights as the starting point for our merger classification training. We hypothesise that the initialisation with the pre-trained \textsc{ImageNet} weights can act as a regulariser. We will thus expect an improvement in the generalisation performance of the classifier to a level that is significantly above the level of just training the merger classifier from random weight initialization. This is one form of transfer learning.

\subsection{Dataset}

To train a classifier, we need a dataset which consists of pairs of images with the corresponding ground truth classifications (\textit{merger} vs \textit{non-interacting}). Our source for the imaging data is the Sloan Digital Sky Survey (\textsc{SDSS}) Data Release 7, where we use the SDSS online image cutout service\footnote{\url{http://skyserver.sdss.org/dr12/en/help/docs/api.aspx}} to get RGB JPEG images of the galaxies of interest. For our ground truth classifications, we are using the crowdsourced labels from the Galaxy Zoo project, where we are interested in the weighted-merger-vote fraction $f_m$. We use the 3003 merger objects from the \cite{Darg2010} merger catalogue, which itself is based on $f_m$ from the GZ data. This catalogue takes all objects with $0.005 < z < 0.1$ and $f_m > 0.4$ and runs them through a second visual confirmation process, using human experts. This yields the 3003 merger objects in the catalogue. As our non-interacting galaxies sample, we choose 10000 GZ galaxies with $f_m < 0.2$ and in the same redshift range in a random draw. During training, we will do stratified sampling from those two sets, so that each mini-batch has the same number of images of merging galaxies and non-interacting galaxies.

\subsection{Experiment}

We are interested in two questions: How does a modern CNN architecture, trained on a merger dataset, compare to previous SOTA merger classifications (main experiment), and how does the classification performance of a CNN with transfer learning compare to a CNN with random initialization (lesion study). We hypothesise transfer learning to be useful especially in cases with small training set sizes, thus we will conduct the lesion study for different training set sizes to investigate the influence of training set size on the utility of transfer learning over random initialization. We will use artificially restricted training sets with training set sizes of $[3000, 1500, 900, 500, 300]$ and test the superiority of a transfer learning approach for each one of them. For more technical details about the training refer to the corresponding parts of the appendix.

\section{Results}

After training the CNN on the training set, we need to evaluate the performance of the classifier on the \textit{test set}, which was never used during training. The trained classifier produces, for each galaxy image, an output $p_m \in [0, 1]$, where a value of $0$ means a classification as a non-interacting galaxy, and a value of $1$ means a classification as a merger system. We chose a threshold of $0.5$ for $p_m \in (0, 1)$ to distinguish between the two categories and get a binary classification.

\begin{table}
	\centering
	\caption{Reporting precision, recall and $F_1$ of our method, with a comparison to previous automatic visual classification methods. Keep in mind that only the recall $r$ can be used as a valid comparison between the methods, as only this quantity is invariant under the different class ratios used for testing by the different authors.}
	\label{tab:previous_methods}
	\begin{tabular}{lccc}
		\hline
		method & recall $r$ & precision $p$ & $F_1$\\
		\hline
		\cite{Goulding2017} & 0.75 & 0.90 & 0.82\\
        \cite{Cotini2013} & 0.8 & 0.8 & 0.8\\
		\cite{Hoyos2011} & 0.92 & 0.29 & 0.44\\
		our method & \textbf{0.96} & \textbf{0.97} & \textbf{0.97}\\
		\hline
	\end{tabular}
\end{table}

After obtaining the automatic classification for each image in the test set, we can quantify the performance of our method. We report precision, recall (or sensitivity) and $F_1$ numbers of our method, and compare it to the performance of previous SOTA methods in table \ref{tab:previous_methods}. Precision $p$, recall $r$ and $F_1$ are defined as follows:

\begin{equation}
\begin{aligned}
p &= \frac{n_{TP}}{n_{TP} + n_{FP}} \\[4pt]
r &= \frac{n_{TP}}{n_{TP} + n_{FN}} \\[7pt]
F_1 &= 2 \frac{p \cdot r}{p + r}
\end{aligned}
\label{eq:precision_recall_F_1}
\end{equation}

Here, $n_{TP}$, $n_{FP}$ and $n_{FN}$ refer to the number of \textit{true positive}, \textit{false positive} and \textit{false negative} classifications respectively.

\begin{figure}
	\includegraphics[width=\columnwidth]{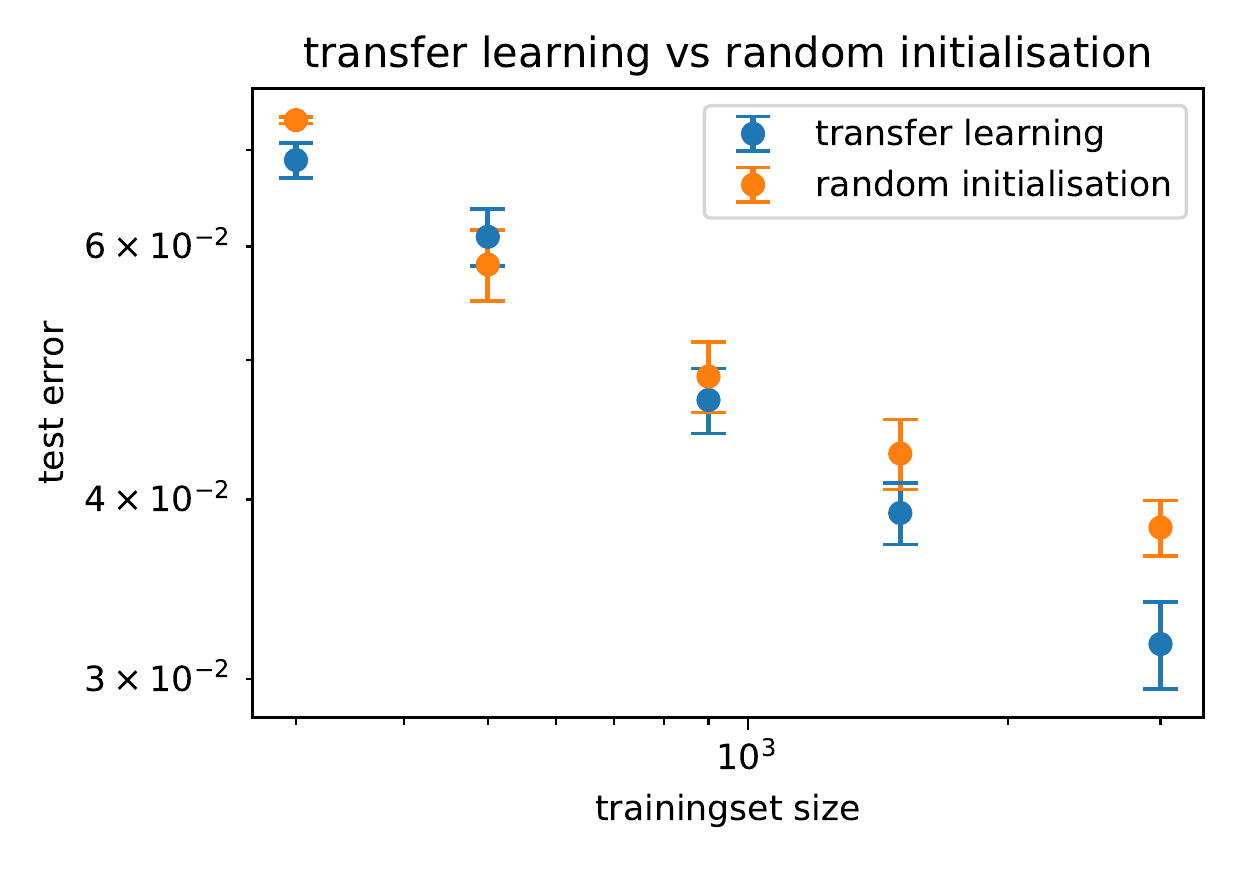}
	\caption{The difference between transfer learning and random initialisation in terms of test error. The experiment was done for five different training set sizes of $[3000, 1500, 900, 500, 300]$. Error bars denote $1\sigma$ (SEM). Generally, we see transfer learning outperforming random initialisation. However, for the individual experiments, we can only statistically confirm that for the training set sizes of 300 and 3000. The combined p-value for all five experiments is $p = 0.02$}
    \label{fig:lesion_study}
\end{figure}

In figure \ref{fig:lesion_study}, we plot the test error rates of the CNN with transfer learning versus the test error rates of the CNN with random initialisation, for the different chosen training set sizes of $[3000, 1500, 900, 500, 300]$.

Using 4-fold cross validation during training (cf. appendix), we obtained four independent samples of the test error, for every single experiment. We can use the one-tailed Welch's t-test \citep{Welch1947} to calculate p-values for the null hypothesis of transfer learning having the same or higher mean test error as random initialisation.

A statistically significant ($p < 0.05$) advantage of transfer learning over random initialisation can be observed for the training set sizes of 300 ($p = 0.05$) and 3000 ($p = 0.03$), a plausible advantage for the training set size of 1500 ($p = 0.12$), and finally inconclusive results for the training set sizes of 900 ($p = 0.32$) and 500 ($p = 0.72$).

If we combine the different experiments using Stouffer's Z-score method \citep{Stouffer1949, Whitlock2005}, we arrive at a combined p-value of $p = 0.02$, which leads us to the conclusion that transfer learning gives a significant advantage over random initialisation.

\section{Quantifying Performance}

\subsection{Histogram and calibration plot}

\begin{figure}
	\includegraphics[width=\columnwidth]{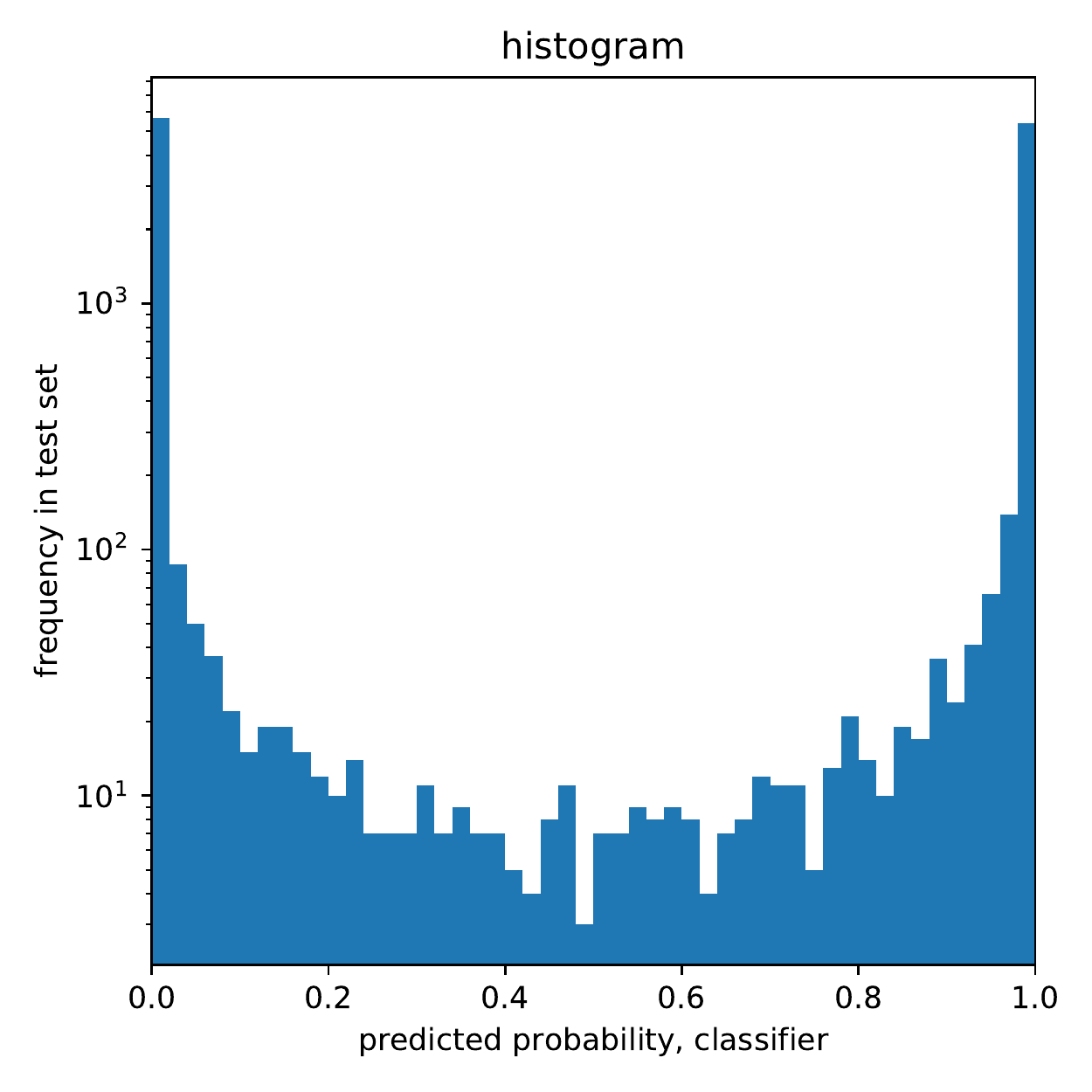}
    \caption{The histogram of $p_{m}$, the output probabilities of the test images being mergers, taken from the CNN output layer. Note that in most cases, the classifier has a high level of certainty and $p_{m}$ lies close to 0 or 1 (i.e. certain classification by the CNN as \textit{non-interacting} or \textit{merger} system, respectively).}
    \label{fig:histogram}
\end{figure}

Classifying all images from the test set with our classifier gives us a  $p_m \in [0, 1]$ for every test set image (A value of $0$ meaning classification as a non-interacting galaxy, and a value of $1$ meaning classification as a merger system). We can immediately plot a histogram of those classifications (cf. figure \ref{fig:histogram}). We can see that the CNN predominantly outputs $p_m$ close to either zero or one, cases with unclear classification are rare.

\begin{figure}
	\includegraphics[width=\columnwidth]{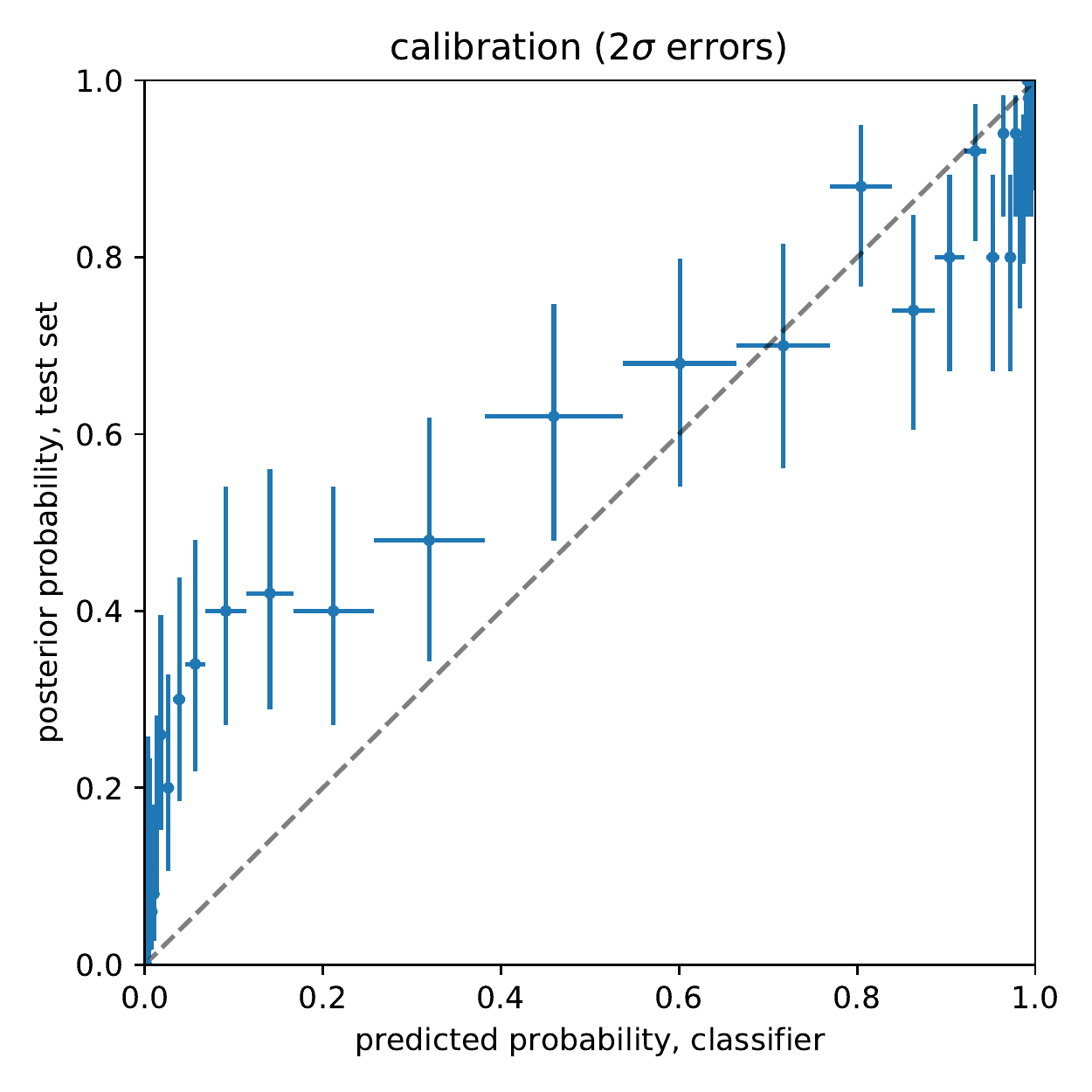}
    \caption{Calibration plot. The error bars in x-direction denote the limits of bins of $p_{m}$, while the error bars in y-direction denote $2\sigma$ CIs for the posterior distribution of the probability parameter $p$ of a Bernoulli trial (using Jeffrey's prior). The calibration is somewhat consistent with a diagonal crossing the origin and of slope 1. However, there's some deviations in the high and low parts of the probability predicted by the classifier. For a recalibrated plot, please refer to figure \ref{fig:calibration_recalibrated}.}
    \label{fig:calibration}
\end{figure}

Let us now focus on all galaxies that were classified into a bin between $p_m$ and $p_m + \Delta p_m$ for some $p_m$, and $\Delta p_m$ small. Then we can, knowing the true classifications in the test set, see if the fraction of true mergers in this bin is indeed close to $p_m$. This is called a calibration plot (cf. figure \ref{fig:calibration}). We can see that the calibration is somewhat close to a diagonal from $(0,0)$ to $(1,1)$, i.e. if we randomly select $N$ galaxies closely around a certain $p_m$, then we can expect roughly $p_m \times N$ true mergers to be in that sample.

\subsection{Recalibration}

\begin{figure}
	\includegraphics[width=\columnwidth]{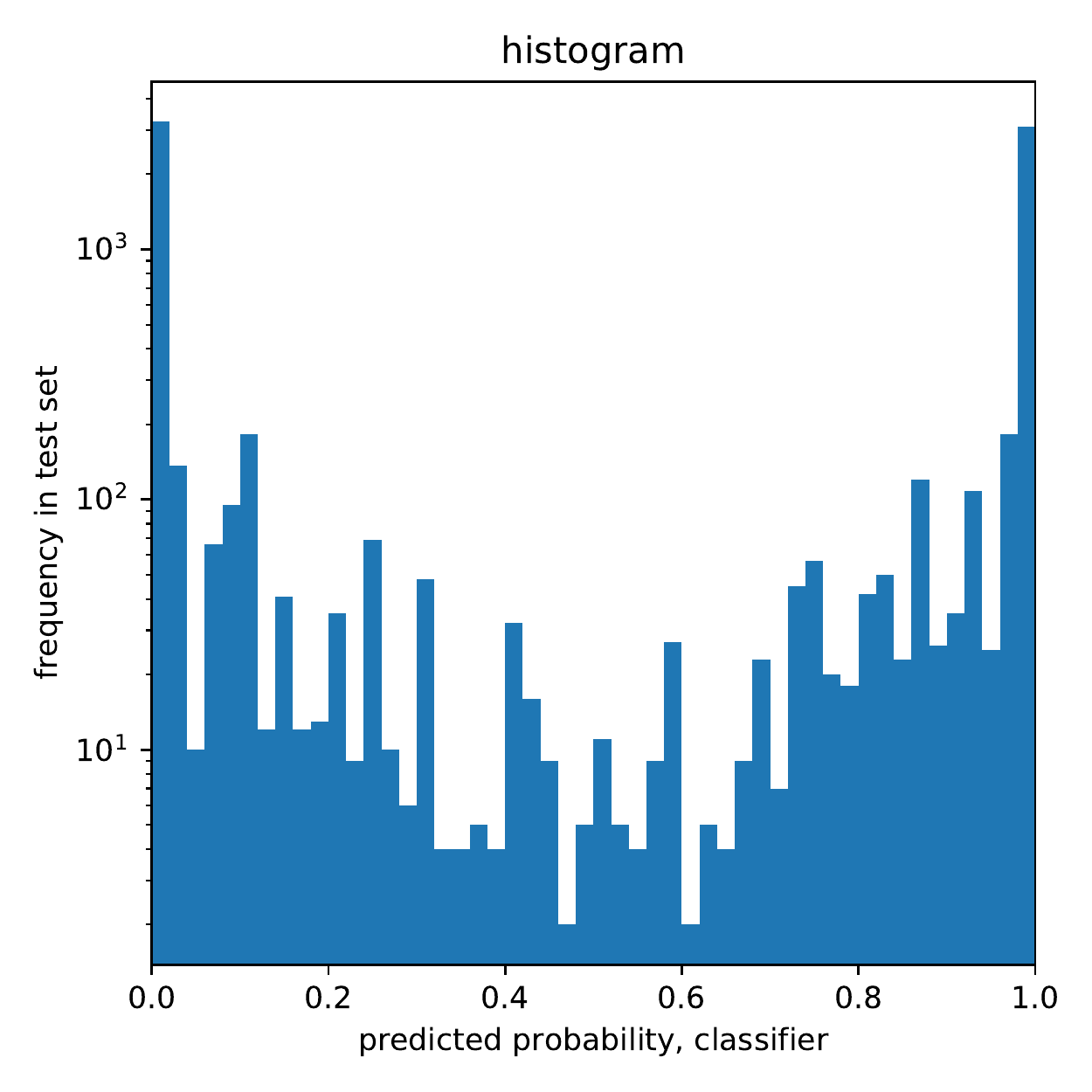}
    \caption{The histogram of $p_{m}$ after recalibration. Note that after recalibration, although the tails of $p_{m} \approx 0$ or $p_{m} \approx 1$ still dominate, there are now more classifications in the middle, i.e. the classifier produces more uncertain classifications (cf. figure \ref{fig:histogram}).}
    \label{fig:histogram_recalibrated}
\end{figure}

\begin{figure}
	\includegraphics[width=\columnwidth]{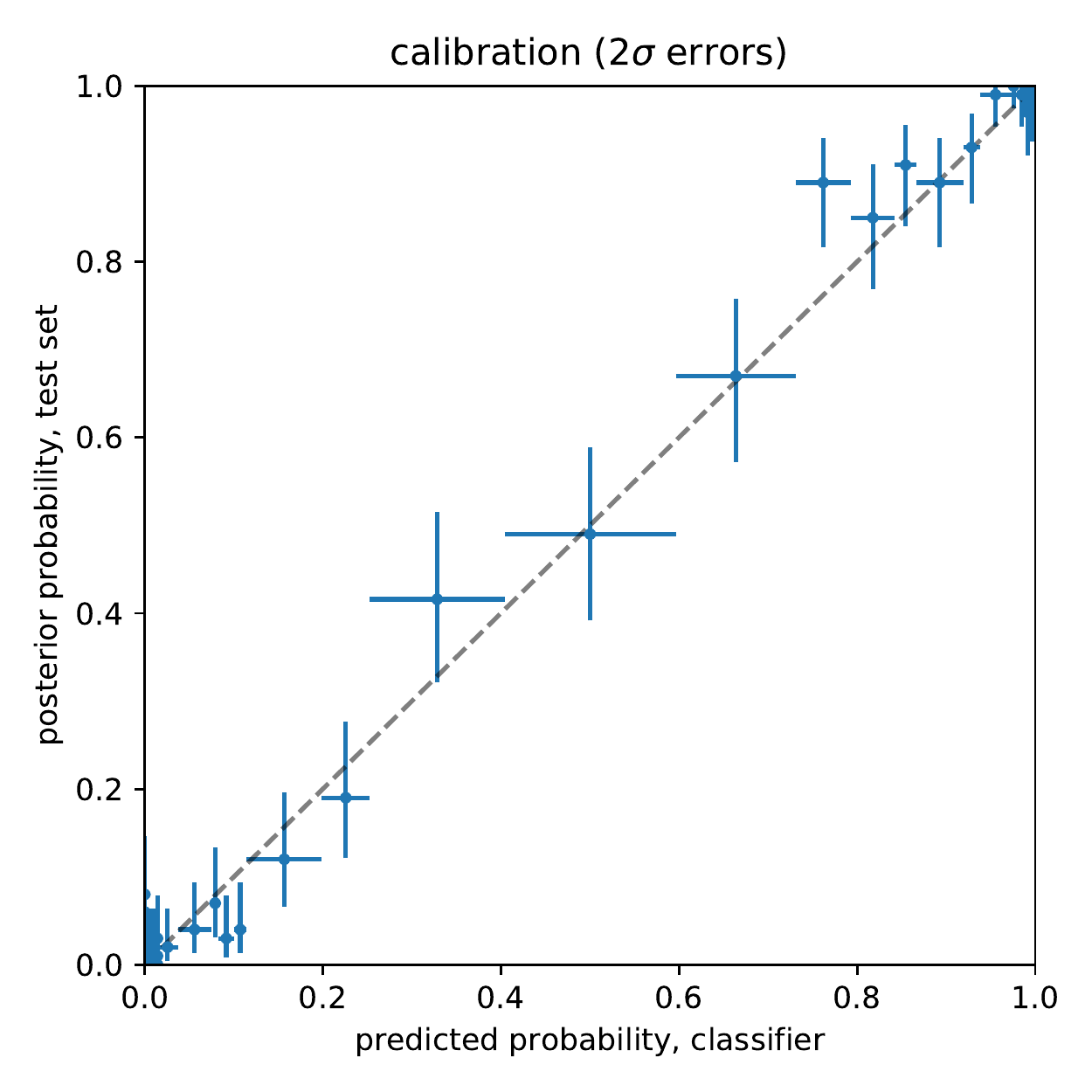}
    \caption{Calibration plot after recalibration. The error bars in x-direction denote the limits of bins of $p_{m}$, while the error bars in y-direction denote $2\sigma$ CIs. The calibration is very consistent with a diagonal crossing the origin and of slope 1, i.e. the classifier is well-calibrated. That means we can interpret $p_m$ as a probability.}
    \label{fig:calibration_recalibrated}
\end{figure}

In order to further improve calibration, i.e. bring the data points in figure \ref{fig:calibration} to a diagonal from $(0,0)$ to $(1,1)$, we employ isotonic regression \citep{Barlow1972, Chakravarti1989}. Isotonic regression is essentially the task of fitting a set of data points optimally with a monotonically non-decreasing model function. Here, we fit a monotonically non-decreasing function $p_{\textrm{true}} \approx c(p_{\textrm{classifier}})$ to the data points in figure \ref{fig:calibration}. This allows us to transform any result directly taken from the classifier $p_{\textrm{classifier}}$, into an approximate true probability $p_{\textrm{true}}$. We used one third of the test set for the recalibration fit, and evaluated the quality on the remaining two thirds of the test set. For the calibration plot and histogram after recalibration, please refer to figures \ref{fig:calibration_recalibrated} and \ref{fig:histogram_recalibrated} respectively.

\subsection{ROC (Receiver Operating Characteristic) curve}

\begin{figure}
	\includegraphics[width=\columnwidth]{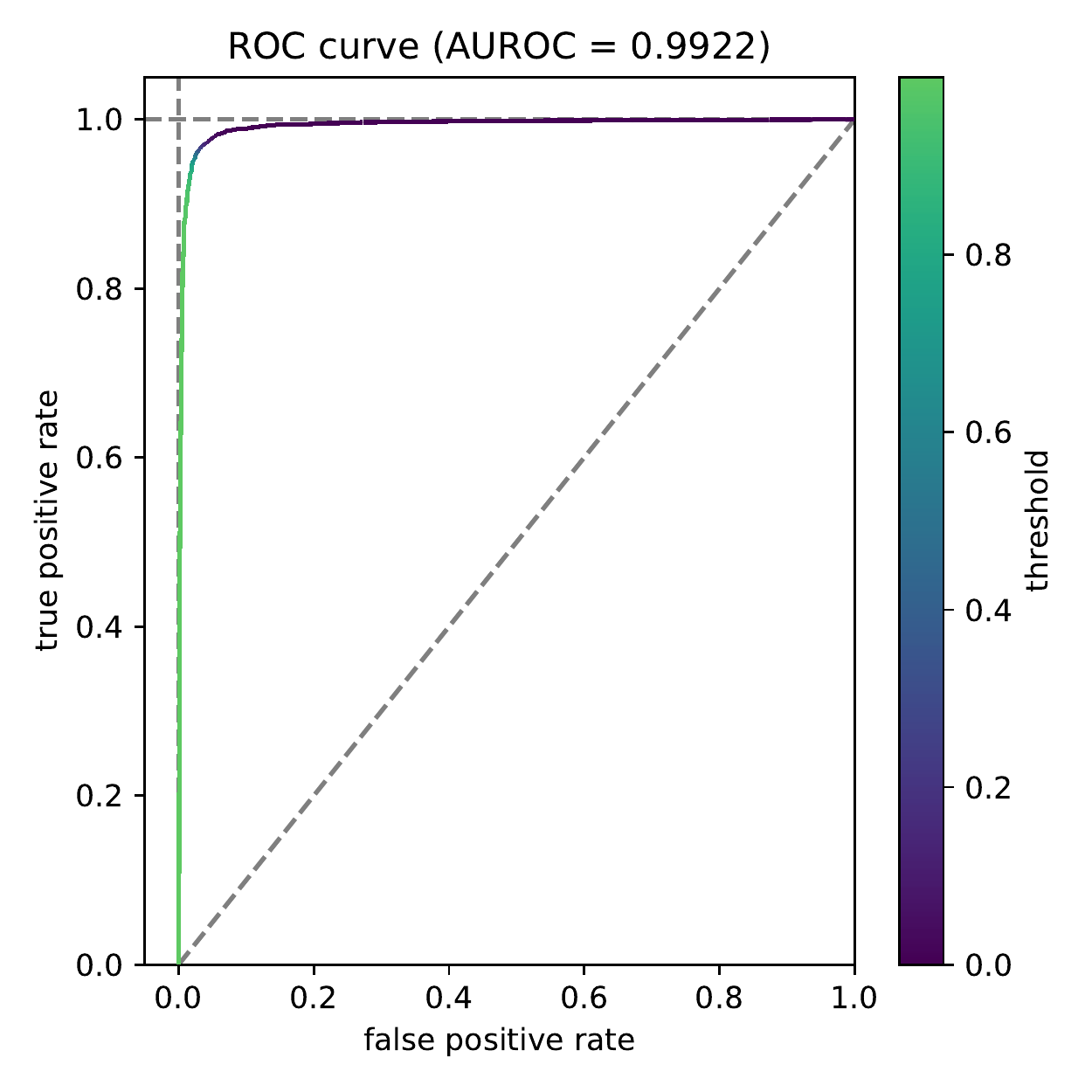}
    \caption{The ROC curve of the classifier. A curve close to the dotted diagonal would correspond to a classifier that just outputs a random $p_m$ for each image. Our ROC curve is consistently and significantly above this diagonal, and reaches an AUROC (integral under the ROC curve) of 0.9922, which is very close to the maximum of 1. This tells us that our classifier performs well, even with high class skew (unbalanced classes) or at any a priori chosen threshold.}
    \label{fig:roc}
\end{figure}

Let us repeat that the output of our classifier is continuous, $p_m \in \left[0, 1\right]$. This means that, if we want a binary classification into either \textit{merger} or \textit{non-interacting} classes, we need to apply thresholding to the obtained $p_m$. The ROC curve is a tool to quantify classifier performance without specifying a particular threshold a priori; we plot true positive rate ($f_{TP} = \frac{n_{TP}}{n_{FN} + n_{TP}}$) versus false positive rate ($f_{FP} = \frac{n_{FP}}{n_{TN} + n_{FP}}$), for all possible thresholds $\in \left[0, 1\right]$. The ROC curve is invariant under changes in class distribution (number of \textit{mergers} versus number of \textit{non-interacting} systems in the test set) \citep{Fawcett2006}. This is useful in our case because we do not know the merger fraction a priori. The AUROC (Area Under ROC), the integral under the ROC curve, results in a single scalar to compare different classifiers; a value close to 1 means a close-to-perfect classifier. The AUROC also quantifies the probability that a randomly chosen \textit{merger} image is classified with a higher $p_m$ than a randomly chosen \textit{non-interacting} system image \citep{Fawcett2006}. In our case we found $\textrm{AUROC} = 0.9922$.

\subsection{Failure modes}

\begin{figure}
	\includegraphics[width=\columnwidth]{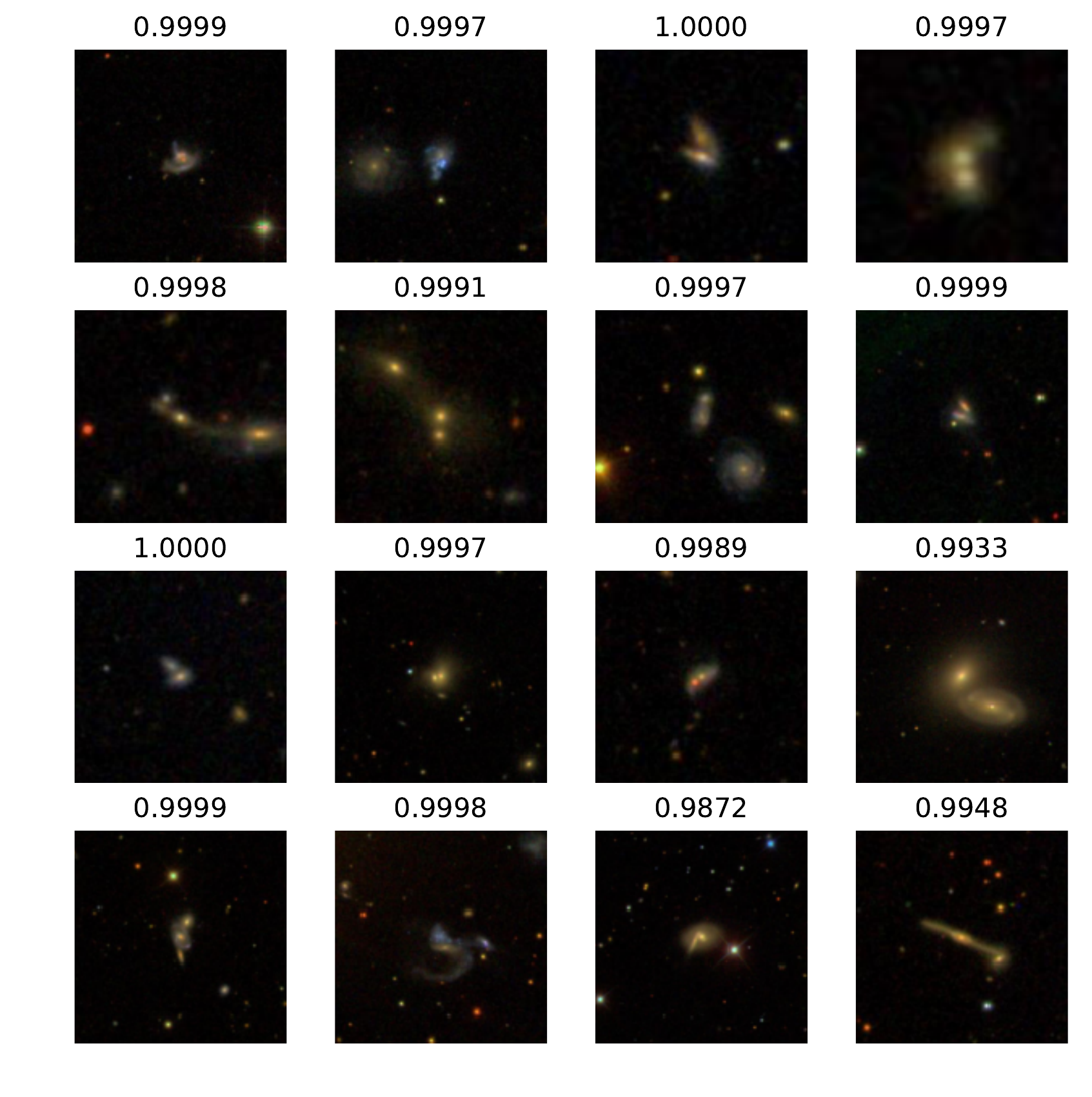}
    \caption{Examples (random draw) for true positive classifications in the test set, including $p_m$. We typically see a high confidence $p_m \approx 1$. The classifier is able to correctly detect a wide variety of different merger morphologies.}
    \label{fig:true_positive_examples}
\end{figure}

\begin{figure}
	\includegraphics[width=\columnwidth]{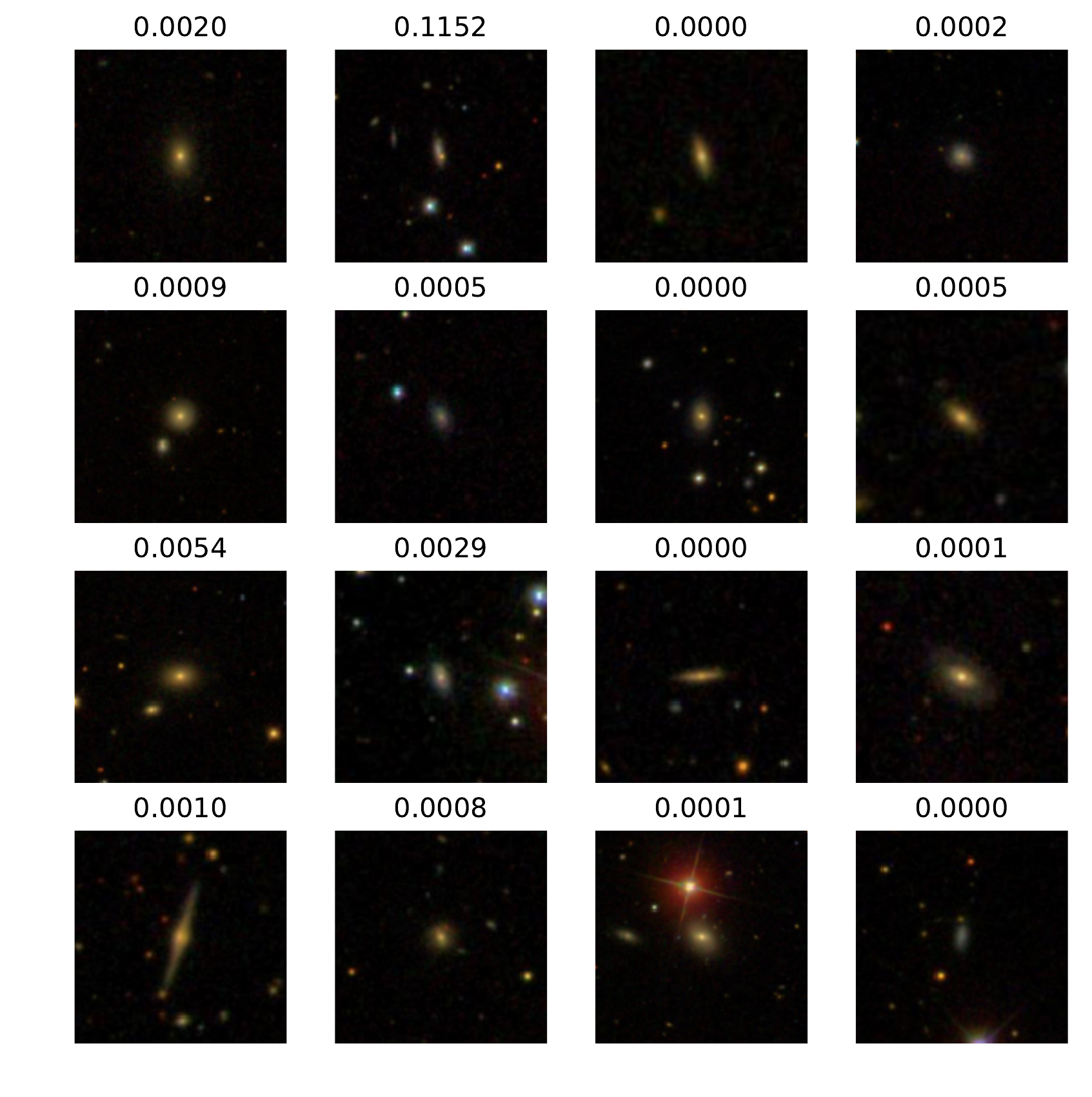}
    \caption{Examples (random draw) for true negative classifications in the test set, including $p_m$. We typically see a high confidence $p_m \approx 0$. The classifier seems to be able to correctly identify star overlaps and still classify the non-interacting galaxy as such.}
    \label{fig:true_negative_examples}
\end{figure}

\begin{figure}
	\includegraphics[width=\columnwidth]{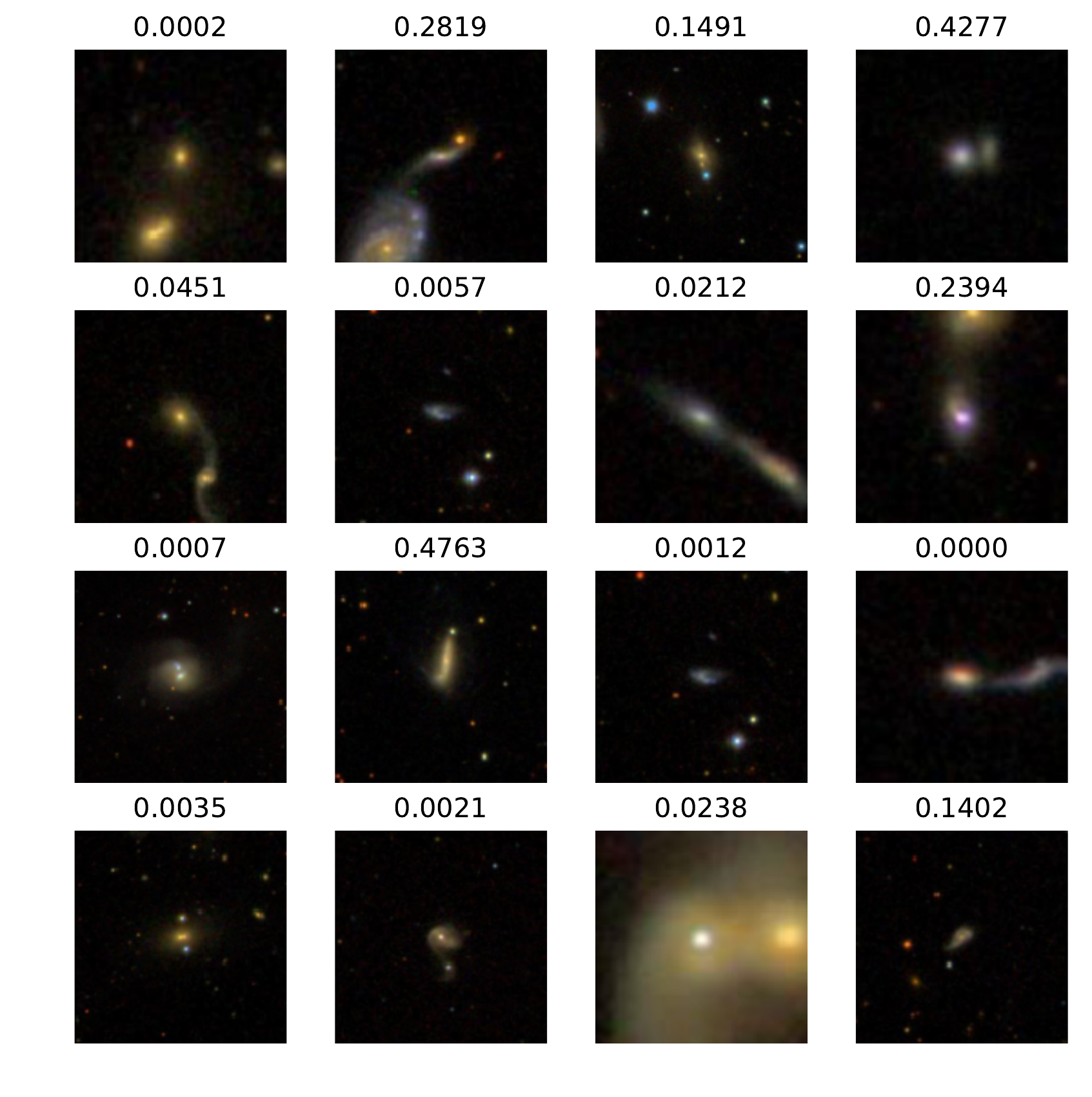}
    \caption{Examples (random draw) for false negative classifications in the test set, including $p_m$. The classifier is less confident for some of the examples, as is expected for a well calibrated classifier. Some false negatives seem to be part of very early or very late stage merger events.}
    \label{fig:false_negative_examples}
\end{figure}

\begin{figure}
	\includegraphics[width=\columnwidth]{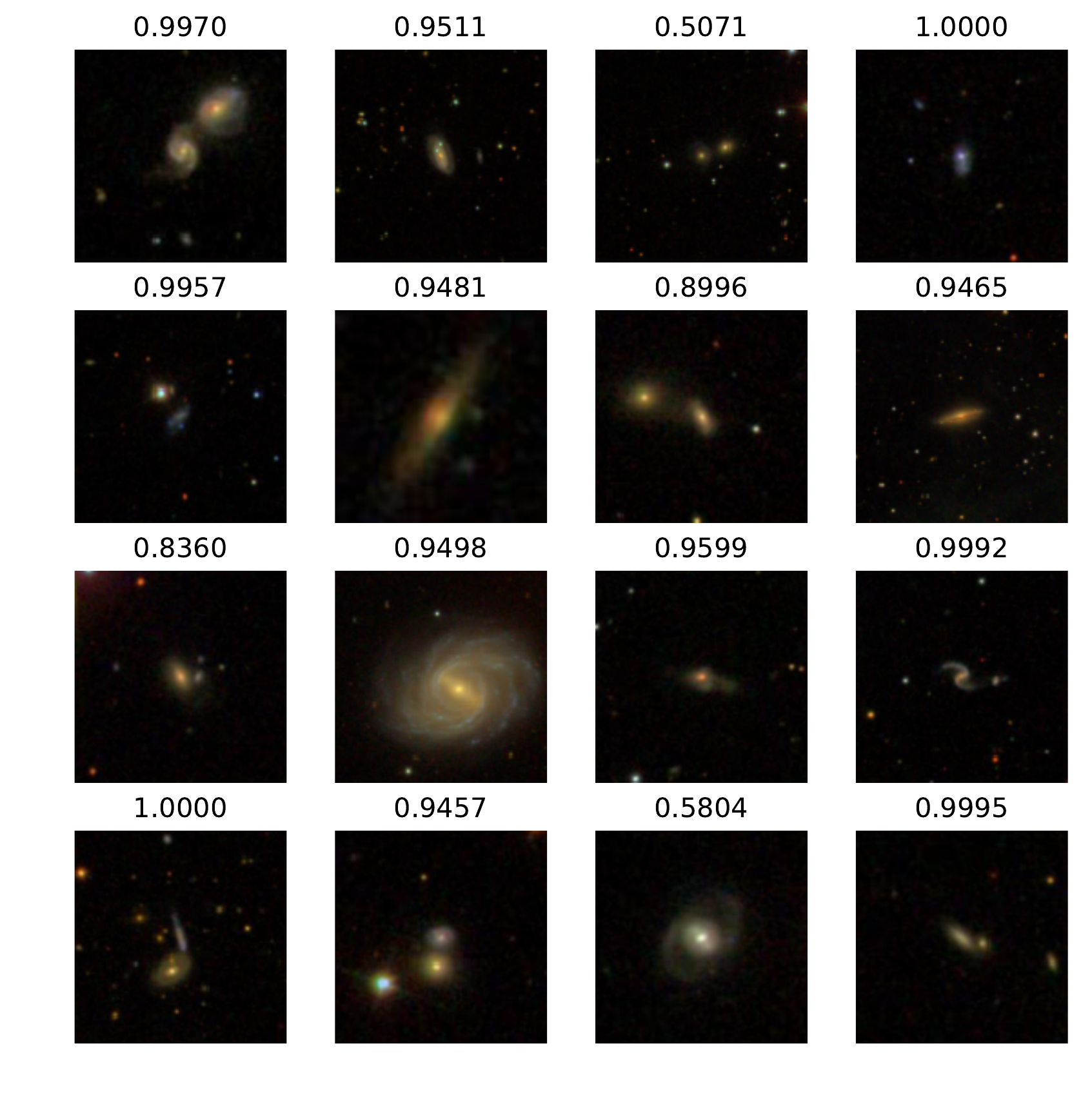}
    \caption{Examples (random draw) for false positive classifications in the test set, including $p_m$. The classifier is less confident for some of the examples, as is expected for a well calibrated classifier. Some false positive examples might be galaxy overlaps. It is also to be expected that a fair number of false positives are actually real mergers, because the \protect\cite{Darg2010} merger catalogue, which we used as the ground truth classifications, was quite conservative in confirming potential mergers, and is certainly not complete.}
    \label{fig:false_positive_examples}
\end{figure}

We provide some example images of true positive/negative and false positive/negative classifications in figures \ref{fig:true_positive_examples}, \ref{fig:true_negative_examples}, \ref{fig:false_negative_examples} and \ref{fig:false_positive_examples} respectively. Overall, the classifier does well with the expected diversity of merger system appearances, and it also seems to be able to correctly identify star overlaps. This level of generality might be quite hard to hand-engineer by adapting a system without feature learning like CAS. On the other hand, a significant part of the misclassifications are images that are also hard to correctly classify for human classifiers, like potential galaxy overlaps and very late and very early stage merger systems.

\section{Properties of the merger sample}

\begin{table*}
	\centering
    \caption{The top and bottom five objects in our merger sample, according to the mean of all $p_m$. This table does not include any objects used during training. Notice that we have four different estimates for $p_m$. This is due to using 4-fold cross validation during training, which gives us four different classifiers.}
	\label{tab:merger_sample}
	\begin{tabular}{lcccccc}
	\hline
		objid &           ra &          dec &  $p_m^{(0)}$ &  $p_m^{(1)}$ &  $p_m^{(2)}$ &  $p_m^{(3)}$ \\
		\hline
		587725552819634248 &  10:56:39.17 &  +67:10:49.0 &    1.000000 &    0.999997 &    0.999994 &    0.999996 \\
		587741816249516036 &  10:37:58.48 &  +22:25:00.0 &    0.999996 &    1.000000 &    0.999981 &    0.999994 \\
		588011124116488195 &  13:15:35.06 &  +62:07:28.6 &    0.999999 &    0.999970 &    1.000000 &    0.999997 \\
		587733080280465515 &  13:25:29.68 &  +53:34:56.3 &    0.999996 &    0.999993 &    1.000000 &    0.999977 \\
		587742014375067679 &  15:07:55.82 &  +17:21:50.9 &    0.999997 &    0.999973 &    0.999974 &    0.999999 \\
    	\vdots &  \vdots &  \vdots & \vdots & \vdots & \vdots & \vdots \\[4pt]
		587745539982032999 &  10:09:15.16 &  +14:49:58.2 &         0.0 &         0.0 &         0.0 &         0.0 \\
		587735348019921176 &  09:28:09.65 &  +10:47:28.5 &         0.0 &         0.0 &         0.0 &         0.0 \\
		587739610240123062 &  12:42:47.60 &  +33:17:15.8 &         0.0 &         0.0 &         0.0 &         0.0 \\
		587731870163140798 &  10:03:47.64 &  +50:40:10.4 &         0.0 &         0.0 &         0.0 &         0.0 \\
		588016891177533614 &  10:45:03.54 &  +39:25:17.0 &         0.0 &         0.0 &         0.0 &         0.0 \\
		\hline
	\end{tabular}
\end{table*}

We created a merger sample by taking all the Galaxy Zoo I objects in the same redshift range as the \cite{Darg2010} merger catalogue, and then obtaining the classification $p_m$ for each galaxy with our classifier. We used the classifier trained on the largest training set with transfer learning. Table \ref{tab:merger_sample} shows the top and bottom five objects according to $p_m$.

Keep in mind that some of the Galaxy Zoo I objects were already in the dataset for training our classifier. However, using 4-fold cross validation during training, there is always at least one classifier, for every single object, that has not been exposed to this object during training. This means we can actually provide a $p_m$ for every object without cheating by doing inference on the training or validation set.

To investigate the properties of this merger sample, we examine the distribution of detected mergers in colour-mass space and determine their stellar mass function.

Please note that here we are using a classifier, trained on a certain training set, for inference on a data set with with similar, but different underlying statistical properties. We do not have any guarantees that this will lead to a sensible merger sample. However, we will argue \textit{ex-post} by comparing the resulting merger catalogue to the \cite{Darg2010} catalogue in terms of a few astrophysical quantities, and finding reasonable agreement between the two catalogues in this regard.

\subsection{Colour-Mass diagram}
\begin{figure*}
	\includegraphics[width=\textwidth]{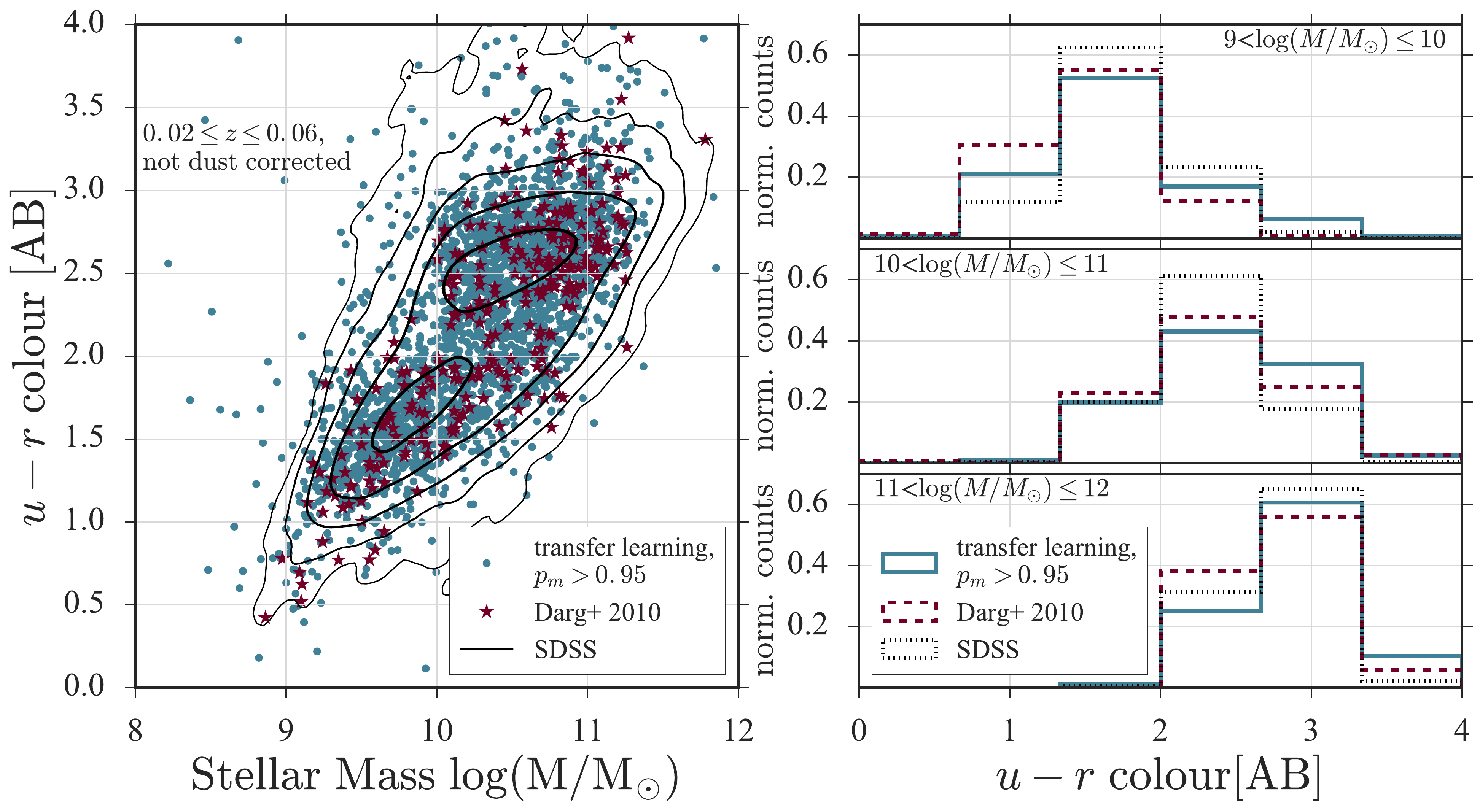}
	\caption{\label{fig:cm}
    Properties of merger galaxies that have been identified with transfer learning. On the left-hand side, we show the distribution of merger galaxies in colour-mass space. The blue filled markers show mergers that have been identified with the CNN ($p_{m} > 0.95$). The red stars show the \protect\cite{Darg2010} sample. We use volume complete samples ($0.02 \leq z \leq 0.06$) and apparent magnitudes that have not been corrected for dust. The black contours show the distribution of all SDSS galaxies in the same volume. On the right-hand side, we show the distribution of mergers and of all galaxies in $u - r$ colour space for three stellar mass bins. Note that here we do not introduce a volume limit, but consider all galaxies within the $0.02 \leq z \leq 0.06$ range. The left-hand side panel illustrates that, similar to the \protect\cite{Darg2010} sources, the transfer learning mergers span the entire colour mass space, from the blue cloud to the red sequence. The right-hand side shows that, compared to the \protect\cite{Darg2010} sample, the CNN mergers show a tendency towards redder colours.}
\end{figure*}

In Fig. \ref{fig:cm} we show the distribution of merger galaxies that have been identified with the CNN ($p_{m} > 0.95$) in colour-mass space. Our (arbitrary) threshold $p_{m} > 0.95$ leaves us with 7980 objects (out of originally 328151). For the colour-mass diagram, shown on the left-hand side, we use a volume complete sample ($0.02 \leq z \leq 0.06$) to avoid a bias due to incompleteness. This selection effect can be avoided if the sample is split into mass bins. On the right-hand side, we thus illustrate the colour distribution of the entire sample in three different stellar mass bins. For comparison, we also show the properties of merger galaxies identified via visual classification \citep{Darg2010} and of all SDSS galaxies within the same volumes. Note, that the apparent magnitudes used here have not been corrected for dust.  

Fig. \ref{fig:cm} illustrates, that major merger galaxies that have been identified via transfer learning lie within the same colour and mass range as visually classified mergers. Both samples span from the blue cloud, across the green valley, to the red sequence \citep{Bell2003,Baldry2004,Faber2007,Martin2007,Schawinski2014}. Comparing the $u - r$ values of CNN and visually classified mergers in more detail shows that, compared to the \cite{Darg2010} sample, our sources tend towards redder colours. However, the Kolmogorov-Smirnov test \citep{eadie1971statistical} shows, that this difference is only significant for the $9 < \log(M/M_{\odot}) \leq 10$ bin ($p$-value = 0.009). Hence only for this bin the two samples are likely to have been drawn from different distributions.

\subsection{Stellar mass functions}
\begin{figure}
\includegraphics[width=\columnwidth]{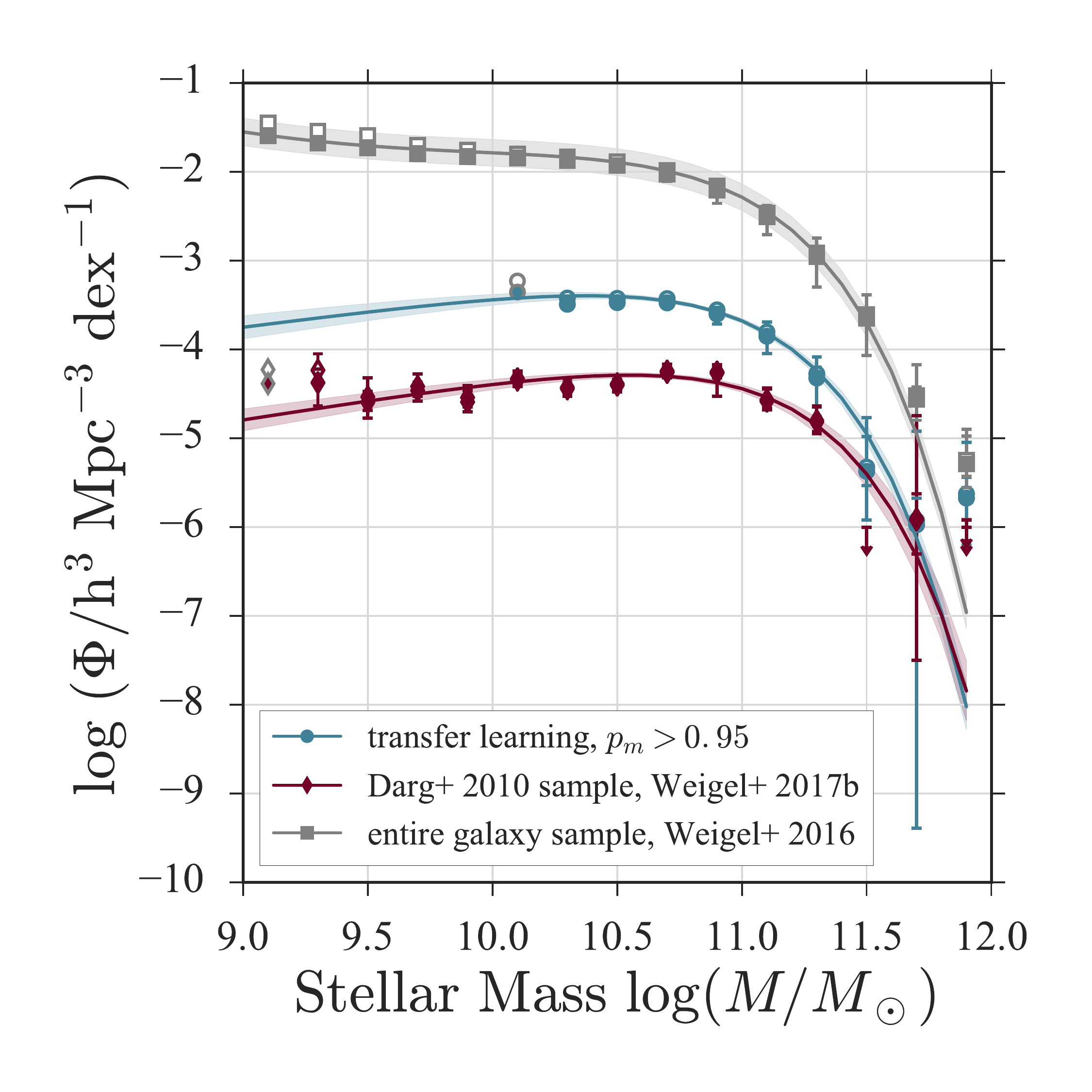}
\caption{\label{fig:smf}
		Stellar mass function of mergers that have been identified with transfer learning in comparison to the mass functions of all galaxies and of visually classified major mergers. In blue, we show the stellar mass function of galaxies with $p_{m} > 0.95$ in the $0.02 < z < 0.06$ range, which we determine by using the method by \protect\cite{Weigel2016}. We compare our results to the mass function of visually selected major mergers (red, \protect\cite{Weigel2017b}) and of all galaxies \protect\cite{Weigel2016} in the same redshift range. Open markers, filled markers, and solid lines show the results of three independent mass function estimators (see text and \protect\cite{Weigel2016} for more details).}
\end{figure}

In addition to the colour-mass diagram, we also determine the stellar mass function of our merger sample. Stellar mass functions are a sophisticated statistical measurement. Their determination includes correcting for selection effects and the resulting shape reflects the details of the true, underlying mass distribution. 

\cite{Weigel2017b} use the \cite{Darg2010} catalogue to determine the stellar mass function of major merger galaxies (mass ratio up to 1:3) in the $0.02 < z < 0.06$ redshift range. They find that the space density of major merger galaxies is well fit by a single Schechter function with $\log (M^{*}/M_{\odot}) = $ \logMstarM, $\log (\Phi^{*}/\rm h^3\ Mpc^{-3}) = $ \logphistarM, and $\alpha = $ \alphaM. We restrict our sample to the same redshift range, select galaxies for which $p_{m} > 0.95$, and use the same method (see \citealt{Weigel2016}) as \cite{Weigel2017b} to determine the stellar mass function of merger galaxies that have been identified with transfer learning.

Fig. \ref{fig:smf} illustrates our results. We show the stellar mass function of merger galaxies that have been identified by the CNN in blue, the mass function of visually classified major mergers by \cite{Weigel2017b} in red, and the mass function of the entire galaxy sample in the same redshift range \cite{Weigel2016} in grey. Open markers ($1/\rm{Vmax}$, \citealt{Schmidt1968}), filled markers (SWML: \citealt{Efstathiou1988}), and solid lines (STY: \citealt{Sandage1979}), illustrate the results of three independent mass function estimators. For the mass function of CNN identified mergers we find the following best-fitting parameters: $\log (M^{*}/M_{\odot}) = $ \logMstarT, $\log (\Phi^{*}/\rm h^3\ Mpc^{-3}) = $ \logphistarT, and $\alpha = $ \alphaT. The shape of the mass function of mergers that have been identified via transfer learning thus resembles the mass function of visually classified major mergers. However, we find a significantly higher space density $\Phi^{*}$.

Using a different cut in terms of $p_{m}$ does not change $\Phi^{*}$. The normalisation, $\Phi^{*}$ does however increase with $p_{m}$. For $p_{\rm m} = 0.90, 0.95, 0.99$ we determine $\log (\Phi^{*}/\rm h^3\ Mpc^{-3}) = $ \logphistarTlow, \logphistarTmed, \logphistarThigh, respectively.

To interpret the $\Phi^{*}$ difference in Fig. \ref{fig:smf}, it is important to be aware of the differences between our and the \cite{Weigel2017b} mass function. \cite{Weigel2017b} restrict their sample to major merger galaxies, i.e. they include a cut in terms of the mass ratio of the merging galaxies. The mass measurements are based on fits to the photometry \citep{Darg2010}. For each merging system, they include the mass of the more massive merging partner, if spectra are available for both merging galaxies. They use the mass of the galaxy for which a spectrum is available, if only one of the merging galaxies has been observed spectroscopically. We do not include a mass ratio cut. Furthermore, we include all spectroscopically observed galaxies with $p_{m} > 0.95$ in our mass function. In contrast to the \cite{Weigel2017b} sample, we thus count systems double, if both merging partners have been observed spectroscopically and both have $p_{m} > 0.95$. Due to these differences in terms of sample selection, the $\Phi^{*}$ offset between our sample and the results by \cite{Weigel2017b} does not directly imply that, within the same volume, we are able to identify more merger galaxies with transfer learning than with a visual classification. 

Fig. \ref{fig:smf} illustrates another subtle difference between our sample of mergers that have been identified with transfer learning and the \cite{Darg2010} sample of visually selected mergers: whereas the \cite{Darg2010} sample is complete to $\log (M/M_{\odot}) = 9$, the completeness of our sample only reaches to $\log (M/M_{\odot}) = 10$. This is due to the difference in terms of colour, which we discussed in the previous section. Tending toward redder colours, our mergers exhibit lower mass-to-light ratios (low luminosity compared to their stellar mass) than the \cite{Darg2010} mergers. Mass-to-light ratios are used to translate a survey's completeness in terms of luminosity into a completeness in terms of stellar mass \citep{Pozzetti2010, Weigel2016}. This conversion is particularly sensitive to the mass-to-light ratios of low mass, low redshift galaxies. The difference in terms of colour, which we illustrated in Fig. \ref{fig:cm} and which was significant for the lowest mass bin, thus directly accounts for the difference in terms of completeness in Fig. \ref{fig:smf}.  

Besides this difference in terms of mass-to-light ratios, Fig. \ref{fig:smf} illustrates the overall consensus between visual and CNN based merger classifications. 

\section{Conclusions}

We have shown that by using state-of-the-art CNNs, we can outperform the previous methods for automatic visual detection of galaxy mergers significantly. We also showed that for our dataset sizes, transfer learning by initialising with e.g. \textsc{ImageNet} weights can lead to a modest improvement in the generalisation power of the trained classifier. A sanity check of our method by creating a merger sample with our method and comparing the properties of this sample to the \cite{Darg2010} catalogue shows agreement in terms of colour-mass distribution and stellar mass function.

Our methods are not specific to merger classification and can be used for the general problem of detecting rare astronomical objects such as gravitational lenses \citep{2016MNRAS.455.1171M}, galaxies with shocked interstellar medium \citep{2016ApJS..224...38A}, AGN ionization echoes \citep{2012MNRAS.420..878K} or ring galaxies \citep{1995ApJS...96...39B}.

We would also like to emphasise the convenient property of our method to produce well-calibrated classifications, i.e. for each image, the classifier calculates a number $p_m \in [0, 1]$, which can be interpreted as a true probability of the classified object being part of a merger system.

Please refer to the \textsc{SpaceML}\footnote{\url{https://space.ml/}} project to access code, full models of the classifiers, and a full table of the GZ I derived merger sample obtained from our classifiers.

\section*{Acknowledgements}
KS and AKW acknowledge support from Swiss National Science Foundation Grants PP00P2\_138979 and PP00P2\_166159 and the ETH Zurich Department of Physics.  CZ and the DS3Lab gratefully acknowledge the support from the Swiss National Science Foundation NRP 75 407540\_167266, IBM Zurich, Mercedes-Benz Research \& Development North America, Oracle Labs, Swisscom, Zurich Insurance, Chinese Scholarship Council, the Department of Computer Science at ETH Zurich, and
the cloud computation resources from Microsoft Azure for Research award program.




\bibliographystyle{mnras}
\bibliography{main} 


\appendix

\newpage
\section{CS Protocol}

\subsection{Main Experiment}
\begin{itemize}
\item {\bf Hypothesis:} CNNs (with transfer learning) are able to outperform
state-of-the-art techniques for merger classification
\item {\bf Proxy:} We measure the quality of our approach in terms of
precision, recall and F-1 score at a classification threshold of $p = 0.5$. We also evaluate the ROC curve and measure the AUROC.

\item {\bf Protocol:} We conduct $K$-fold cross validation in the following way: For each of the $K$ iterations, we use one fold as the validation set, one fold as the test set, and the rest of the folds as the training set. First, we replace the last fully connected layers, trained on the the original \textsc{ImageNet} dataset, with two fully connected layers (random initialisation) with only two outputs (corresponding to our two-class classification task). We train just those FC weights for 40 epochs with the rest of the layers frozen. We then use
SGD with momentum $0.9$ and a learning rate of $0.000015$ and use the validation set accuracy for early stopping. We report the quality scores on the test set from the cross validation loop. This leaves us with $K$ samples for each one of our quality scores.
\item {\bf Expected Result:} The classification performance of CNNs with transfer learning dominates state-of-the-art methods, according to the chosen metrics.
\end{itemize}

\subsection{Lesion Study: The Impact of Transfer Learning}
\begin{itemize}
\item {\bf Hypothesis:} Transfer learning outperforms
deep learning with random initialisation.
\item {\bf Proxy:} We compare the different outcomes in term of test error.
\item {\bf Protocol:}
We first generate subsets of the
dataset with sizes of 10\%, 17\%, 30\%, 50\% and 100\%
of the full dataset. For each of these datasets,
we run $K$-fold
cross validation, once using transfer learning and once using random initialisation. This results in $5 \cdot 2 = 10$ independent
$K$-fold cross validation experiments.
We use the same protocol for transfer learning as in the main experiment above. For deep learning  with random initialisation, we first randomly initialise the weights of the network and then use SGD with momentum $0.9$ and a learning rate of $0.000015$ for training. We use the validation set accuracy for early stopping. We report the quality scores on the test set from the cross validation loop. This leaves us with $K$ samples of the test error, for each one of the ten cross validation runs.
\item {\bf Expected Result:} Comparing the quality scores for the two approaches, one should observe
that the quality of transfer learning is better compared to random initialisation, especially for small training set sizes.
\end{itemize}

We choose $K$ by taking into account both available data and available computational resources. We end up choosing $K=4$.

During training, we re-balance the data using stratified sampling for each mini-batch; i.e. each mini-batch contains the same number of images of \textit{merger} and \textit{non-interacting} systems.

\clearpage
\newpage

\bsp	
\label{lastpage}
\end{document}